\documentclass[aps,prb,twocolumn,superscriptaddress]{revtex4}
\usepackage{graphicx}
\usepackage{natbib}

\newcommand{\ket}[1]{| #1 \rangle}
\newcommand{\bra}[1]{\langle #1 |}

\newcommand{\ex}[1]{\langle #1 \rangle}

\newcommand{\beq}{\begin{eqnarray}}
\newcommand{\eeq}{\end{eqnarray}}

\begin{document}

\title {Detecting quantum-coherent nanomechanical oscillations using the current-noise spectrum of a double quantum dot}
\author{Neill Lambert}
\affiliation{Advanced Science
        Institute,
     The Institute of Physical and Chemical Research (RIKEN), Saitama 351-0198, Japan}
 \email{nwlambert@riken.jp}
\author{Franco Nori}
\affiliation{Advanced Science
        Institute,
     The Institute of Physical and Chemical Research (RIKEN), Saitama 351-0198, Japan}
\affiliation{Center for Theoretical Physics, Physics Department,
Applied Physics Program, Center for the Study of Complex Systems,
The University of Michigan, Ann Arbor, Michigan, 48109-1040, USA}
\begin{abstract}
We consider a nanomechanical resonator coupled to a double quantum
dot.  We demonstrate how the finite-frequency current-noise spectrum
through the double quantum dot can be used to distinguish classical
and quantum behavior in the nearby nano-electromechanical resonator.
We also show how the full frequency current-noise spectrum gives
important information on the combined double quantum dot-resonator
energy spectrum. Finally, we point out regimes where the quantum
state of the resonator becomes squeezed, and also examine the
cross-correlated electron-phonon current-noise.

\end{abstract}

\maketitle

\section{Introduction}

The transduction of mechanical motion of resonators and
cantilevers~\cite{Craig, Cho, Roukes1, Treutlein07,Franco1,blencowe}
has become increasingly important with the observation of motion on
the nanometer-scale. In particular, when the ground state energy of
the resonant mode of the mechanical system becomes larger than the
thermal background temperature, a quantized state involving millions
of molecules would materialize.  As nanoelectromechanical systems
(NEMS) reach this regime it becomes increasingly feasible, and
desirable, to transduce their motion by coupling it to a quantum
degree of freedom, like spin\cite{Lambert08},
charge\cite{Rodrigues}, or flux. However, the challenge of finding
an appropriate measuring apparatus, one whose back-action would not
destroy the fragile quantum state, has not been overcome, even if
such devices could be cooled below the quantum limit \cite{Schwab2,
Ouyang}.

Here we propose using a quantized two-level `mesoscopic transport'
degree of freedom, or `transport qubit', as a transducer of quanta
exchange, and to identify signatures of quantum coherent {\em
coupled} phenomena between the mechanical resonator and the
transport qubit. If successfully observed, this would validate the
existence of a quantized mechanical state.  Here we focus on a
capacitively-coupled double quantum dot realization for the
transport qubit. However, our analysis applies to several other
possible devices, such as superconducting single-electron
transistors (SSET) \cite{Schwab2} and suspended double quantum dots
\cite{Blich}, which will be described later.

\subsection{Probing mesoscopic transport}

It is important to note that the types of experimental measurement
that can be made on mesoscopic transport systems are limited; we can
measure the average rate of particles leaving the system (current),
the correlation between these currents at long times (the
zero-frequency noise), and the full Fourier transform of these
correlations (full frequency noise). Over the last few years, the
zero-frequency noise has been used with great success to
experimentally verify coherent quantum behavior (see, e.g.,
Ref.~[\onlinecite{kiesslich}]), and may in the future serve as an
entanglement measure \cite{Lambert07}, and perhaps even aid in
realizing a solid state test of Bell's inequalities~\cite{Emary04}.
The {\em full frequency noise spectrum}, often more difficult to
measure in practice, is appealing because it contains information
about the full dynamics of the
 system: it reveals both coherent dynamics stemming from the system Hamiltonian
$H$, and incoherent dynamics from the environment. This makes it a
powerful tool for probing solid-state quantum systems.

\subsection{Summary of our results}

Our main result here is that we show how the coupled quantum
coherent behavior, e.g., Rabi oscillations, and the low-energy part
of the coupled double quantum dot-resonator spectrum, can be
observed
 as resonances in the {\em full frequency current} noise spectrum. We also analyze the
effects of temperature and decoherence on this signal, and show how
the transition to the classical regime can be monitored using our
approach.

\begin{figure}[htp]
\centering
\includegraphics[width=3.3in]{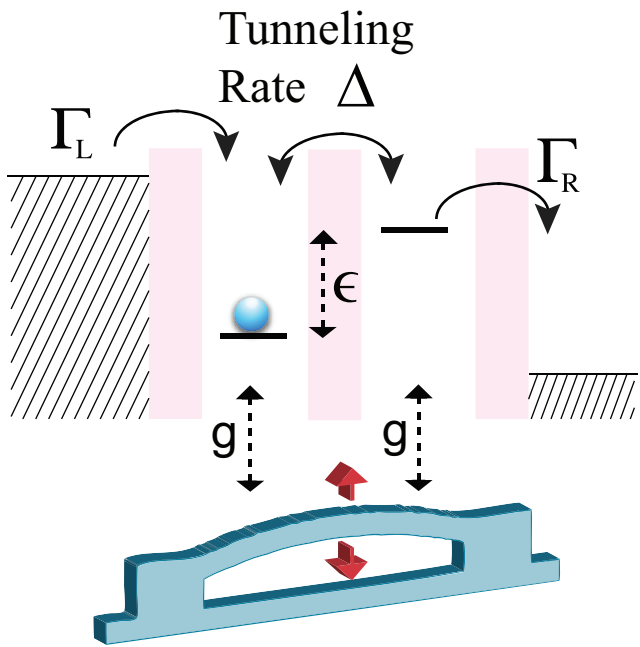}
\vspace{0.0cm} \caption{(Color online) Schematic diagram of a double
quantum dot transport qubit (pink) coupled to a mechanical resonator
(in blue).
 We assume a capacitive coupling $g$ between the position of the resonator and the
 electron charge state of the electron in the dot.  The double quantum dot is attached to two electron reservoirs in the Coulomb blockade regime,
 with tunneling rates $\Gamma_L$ and $\Gamma_R$.  There is a coherent tunneling rate $\Delta$ between the two charge states, and a tunable energy gap $\epsilon$.  We assume that the mechanical resonator is already cooled
 to near the quantum limit \cite{Schwab2, Ouyang}.  The parameter $g$ is the coupling strength between the double quantum dot and the mechanical resonator, whose fundamental frequency is $\omega_b$.  In the schematic diagram, the placement of the components is purely
illustrative. Our model is also applicable to circuit QED systems,
where
 coherent energy exchange between a resonator and a two-level system has recently been observed (see, e.g., Ref.~[\onlinecite{Deppe}]).}
\end{figure}

We now proceed as follows:  we first define a general model for a
transport `qubit' coupled to the quantized fundamental mechanical
mode of a nano-electromechanical resonator. This is a well-studied
model in various forms, and has been used to illustrate, e.g., boson
steering and micromaser effects \cite{Lambert03,Rodrigues,Bennet07}.
Following this, we explain why {\em current-noise measurements} can
contain signatures of {\em quantum} coherent behavior. We illustrate
this with results from our master equation model for two different
parameter regimes.
  We also identify signatures of
quantum state squeezing~\cite{Franco2} of the resonator
, and we calculate the correlation between tunneling events in the
transport qubit and phonons leaving the mechanical resonator.
Finally, we discuss other possible experimental realizations, such
as suspended
 double quantum dots
\cite{Lambert03,Blich,Brandes}, spin states coupled to a magnetized
resonator~\cite{Lambert08}, and capacitively-coupled superconducting
single electron transistors~\cite{Schwab2}.

\section{Model: Transport qubit coupled to a mechanical mode}

The basic Hamiltonian for a `transport qubit' coupled to a quantized
mechanical resonator is as follows, \beq H=\epsilon \sigma_z +
\Delta \sigma_x + g\sigma_z (a+a^{\dagger}) +
\omega_ba^{\dagger}a.\label{1}\eeq Here $\omega_b$ is the
fundamental frequency of the resonator, $\epsilon$ is the energy
gap, or splitting, of the transport qubit states, and $\Delta$ is
the coherent tunneling rate between the two qubit states.  The
bosonic operators $a,a^{\dagger}$ destroy and create excitations in
the resonator. The quasi-spin basis describes the two possible
states in our transport qubit, and we assume that the transport
process enters and leaves through the eigenstates of $\sigma_z$. For
example, for our case of a double quantum dot in the Coulomb
blockade regime, $\sigma_z=\ket{L}\bra{L}-\ket{R}\bra{R}$, where $L$
and $R$ represent an excess electron (N+1 total electrons) in the
left or right dot, and the state $\ket{0}$ represents the empty
state (N total electrons). Note that the excess electron in the
double dot is well separated in energy from the other electrons due
to Coulomb blockade. Alternatively, the superconducting
single-electron transistor can be defined by
$\sigma_z=\ket{2}\bra{2}-\ket{0}\bra{0}$, representing the
superposition of charge states on the island.  Even though
superconducting single-electron transistors are three terminal
devices, in certain regimes the model is equivalent to a double
quantum dot \cite{tobias04} (see below). The spin-blockade
 case would involve a direct coupling, via the magnetization of
the resonator \cite{Lambert08}, to the electron spin
$\sigma_z=\ket{\uparrow}\bra{\uparrow}-\ket{\downarrow}\bra{\downarrow}$.
  Hereafter we retain the double quantum dot
basis, $\{\ket{L},\ket{R}\}$.

\subsection{Master Equation}
Transport, in all these cases, is in non-equilibrium (left to
right), with a large bias applied to the device, and the current
measurement monitors the electrons/particles leaving the device into
the right lead/reservoir (here we neglect displacement-current
contributions). The full equation of motion (master equation) for
this system is described by a super-operator Liouvillian $L$ that
defines the transport of particles through the ``qubit" (under the
Born-Markov approximation), bath damping and temperature terms for
the resonator,
\begin{eqnarray}\label{master1}
\frac{d}{dt}\rho(t) &=& L[\rho(t)]= -i[H,\rho(t)] +L_0[\rho(t)]\\
L_0[\rho(t)]&=&-\frac{\Gamma_L}{2}\left[s_L s_L^\dagger \rho(t) -
2s_L^\dagger \rho(t)s_L + \rho(t)s_L s_L^\dagger\right]\nonumber\\
&-&\frac{\Gamma_R}{2}\left[s_R^\dagger s_R \rho(t) - 2s_R
\rho(t)s_R^\dagger + \rho(t)s_R^\dagger s_R\right]\nonumber\\
&+&\frac{\gamma_b}{2}\left[ - a^\dagger a \rho+2a \rho a^\dagger -
\rho
a^\dagger a\right]\nonumber\\
&+&\bar{n}\gamma_b\left[- a^\dagger a \rho+a \rho
a^\dagger+a^{\dagger} \rho a  - \rho a^\dagger a\right]\nonumber
\end{eqnarray}
where \beq s_L=\ket{0}\bra{L},\,\,\,\,\,\, s_L^{\dagger}=\ket{L}\bra{0},\\
 s_R=\ket{0}\bra{R},\,\,\,\,\,\, s_R^{\dagger}=\ket{R}\bra{0},\\
 \bar{n}= e^{-\hbar \omega_b/kT}/(1-e^{-\hbar \omega_b/kT}),
 \eeq
$\Gamma_L$ and $\Gamma_R$ are the left/right tunneling rates,
$\gamma_b$ is the decay rate of vibrational quanta into the
resonator thermal bath, and $T$ is the temperature of the resonator
thermal bath (hereafter we set $k=\hbar=1$).  $\rho(t)$ is the
density matrix describing the state of the resonator and the qubit.

\subsection{Current-noise power}

We derive the counting statistics of Eq.~[\ref{master1}] using a
generating-function approach (Appendix A). Using these equations we
can calculate the current-noise power~\cite{Blanter00} \beq
S(\omega)_{i,j}&\equiv& \int_{-\infty}^{\infty}d\tau e^{i\omega
\tau}
\left[\ex{\delta{I_i(t+\tau)},\delta{I_j(t)}}\right]_{t\rightarrow
\infty}\eeq where $\delta I_i(t)$ are the current fluctuations, and
$t\rightarrow \infty$ implies the fluctuations are around the
steady-state expectation values. This formalism describes:

(i) particle transport through our effective `qubit' ($i=j=e$,
electron or particle current),

(ii) the statistics of bunching `vibrational phonons' lost to the
background thermal bath of the resonator ($i=j=b$, where $I_b$ is an
effective `bosonic' current),

(iii) correlations between electron and phonon events ($i\neq j$,
$i=e$, $j=b$).

The electron current is defined by the operator \beq \hat{I}_e =
\Gamma_R\; s_R\; \rho(t)\;s_R^\dagger.\eeq Similarly, the
vibrational phonon current is defined by the operator \beq \hat{I}_b
= \gamma_b\; a\; \rho(t)\; a^{\dagger}. \eeq Even though such
`phonon statistics' are typically not experimentally accessible, we
include them here because of the connections of our model to circuit
QED systems \cite{Deppe,you05,You03,You032, Mar05}, where the {\em
photon} statistics can be probed with incident microwave fields and
the state of the pseudo-spin (qubit) by suitable detectors.  Such a
system would also be suitable for observing the cross-correlation
measurements we present later. Also, it is interesting to point out
that in some sense the vibrational mode of the resonator itself can
be thought of as an ``acoustic phonon" with low frequency and long
wavelength. Thus in this manuscript, for brevity we often refer to
the ``vibrational quanta of the fundamental mode of the resonator''
as phonons.

\section{Poles in the current-noise frequency spectrum}

To understand why the current-noise spectrum contains direct
signatures of coherent quantum behavior, we must consider its
dependence on the superoperator $L$.  As discussed by Emary et al
\cite{Emary08}, and Flindt et al \cite{Flindt08} the eigenvalues,
$\alpha_k$, of the superoperator $L$, [e.g. Eq. (2)], consist of
imaginary ``coherent" quantum mechanical level-splitting terms,
originating from $H$, and of real
 ``incoherent" terms, originating from background thermal baths and
non-equilibrium tunneling events.  This can be seen by expanding the
density matrix $\rho$ of the coupled system across the eigenstates
of $H$, $\rho=\sum_{i,j} c_{i,j} \ket{i}\bra{j}$, then the
Liouvillian $L$ acts as \beq L[\rho]&=&-i H\rho
+i\rho H + L_0[\rho]\\
&=&-i(\lambda_i - \lambda_j)\rho + L_0[\rho]\nonumber \eeq As
mentioned above, since all the operators in $L_0$ are real, the
eigenvalues of $L$ will consist of imaginary terms due to energy
level splitting \beq \delta E=(\lambda_{i}-\lambda_j)\eeq and real
terms from operators in $L_0$.

In certain conditions~\cite{Emary08}, the current-noise power can be
expanded in terms of eigenvalues $\alpha_k$ of $L$ and the
coefficients $c_k$ of the matrix $(V^{-1}\hat{I}_e V)_{kk}$, where
$\hat{I}_e$ is the current operator discussed earlier, and $V$ are
the eigenvectors of $L$, so that \beq
S(\omega)=1-2\sum_{k=1}^{N_v}\frac{c_k\alpha_k}{\omega^2
+\alpha_k^2}.\eeq Here, $N_v$ is the dimension of the superoperator
$L$.   If the incoherent terms, those outside the commutator in the
Liouvillian $L$ [e.g. in Eq. (2)], are much bigger than the coherent
energy level splitting $\delta E$ (e.g., $\Gamma_{L,R}, \gamma \gg
\delta E$), then the eigenvalues of the Liouvillian are real, and
the quantum noise is a slowly-varying function of frequency. If,
however, the coherent terms in Eq.~(2) dominate,
  then there exist poles in the current-noise spectrum around the
absolute value of the energy level splitting ~\cite{Emary08}\beq
\omega= |\delta E + i\Gamma + O(\Gamma/\delta E)|,\eeq giving rise
to the resonant features we seek.

\section{Observation of quantum coherence}

To illustrate how to observe quantum signatures, we now investigate
the above model, Eq.~(2), in two regimes: (1) an effective
Jaynes-Cummings regime (when the level splitting matches the
resonator frequency $2\Delta=\omega_b$), (2) and an off-resonance
regime (where $2\Delta \neq \omega_b$).  Later on we will look at
the zero-frequency noise, and make a comparison to a recent
experiment which measured the zero-frequency noise of a double
quantum dot in contact with a many-mode phonon bath.

Hereafter, all results are calculated using the master equation and
noise formalism described above, with a bosonic cut-off appropriate
for the parameter regimes being discussed.   We also discuss, where
appropriate, the dynamics of an effective pure-state, to understand
how the energy spectrum of $H$ contributes to the spectral structure
of the noise.

\begin{figure}[htp]
\centering
\vspace{0.0cm}
\includegraphics[width=2.7in]{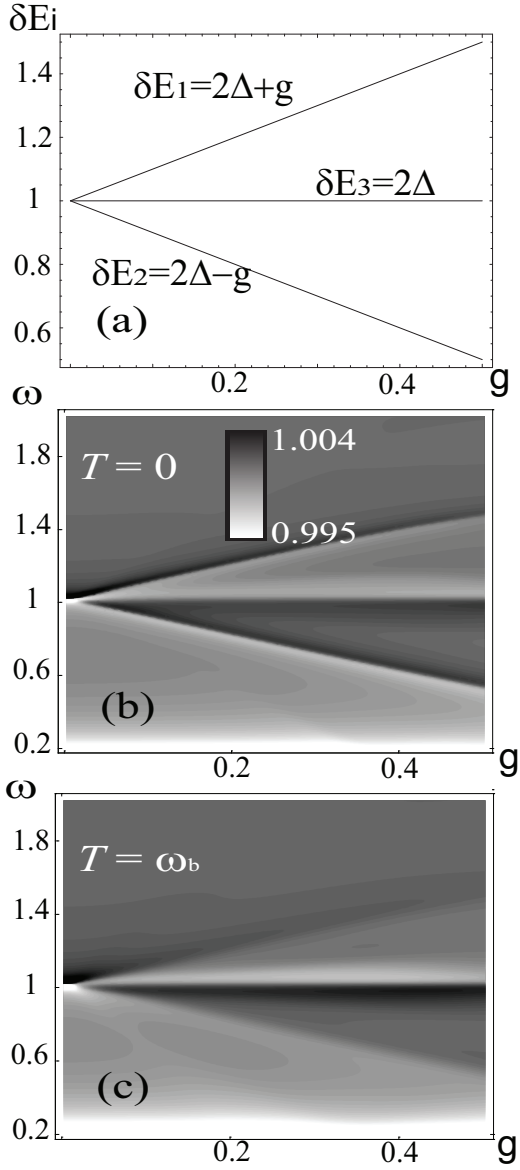}
\vspace{0.0cm} \caption{(a) shows the three important energy gaps,
($\delta E_i$, $i=1,2,3$), in the low-level energy spectrum of the
 Jaynes-Cummings Hamiltonian (see text).  The parameter $g$ is the
coupling between
the electron and the resonator.  
(b, c) show, as contour plots, how these gaps can be observed in the
current-noise frequency spectrum $S(\omega)_{e,e}/2eI_e$, for
$\omega_b=1$, $\Gamma_L=\Gamma_R=0.01$, $\Delta=0.5$,
$\gamma_b=0.05$ and $T=0,\omega_b$, for (b), (c), respectively.
 The energy gaps shown in (a) are clearly visible as three resonances in (b) and (c). The horizontal line corresponds to the physical
 process of an electron tunneling without exchanging quanta with the resonator.  The top and bottom resonant lines are proportional to the coupling $g$, and thus represent the physical
 process of the electron coherently emitting a phonon into the resonator.
 As the temperature is increased, the visibility of the two `Rabi peaks', which are signatures of coherent quantum behavior of the resonator, decreases.}\label{RWA1}
\end{figure}

\begin{figure}[htp]
\centering \vspace{0.0cm}
\includegraphics[width=3.3in]{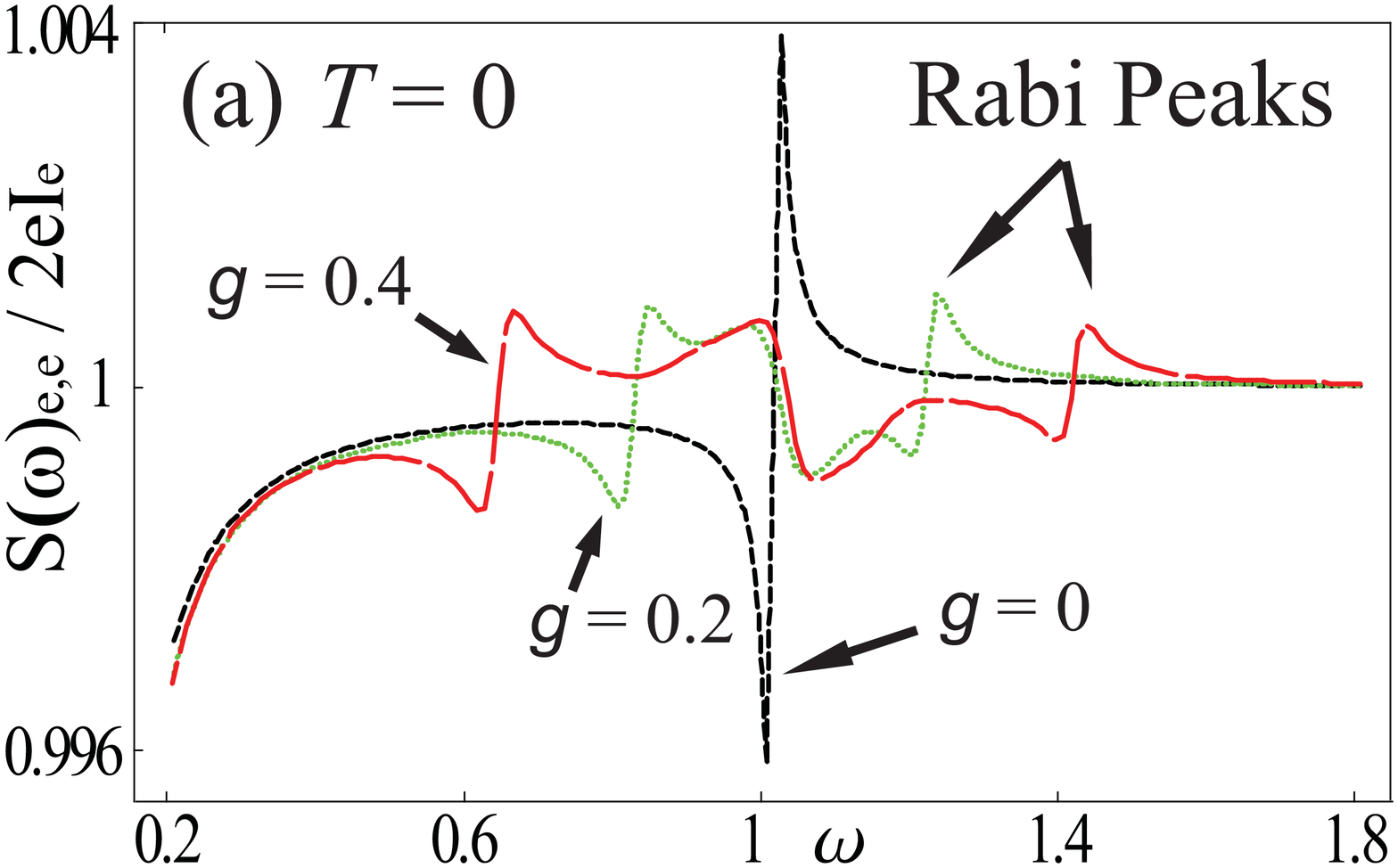}
\vspace{0.0cm}
\includegraphics[width=3.3in]{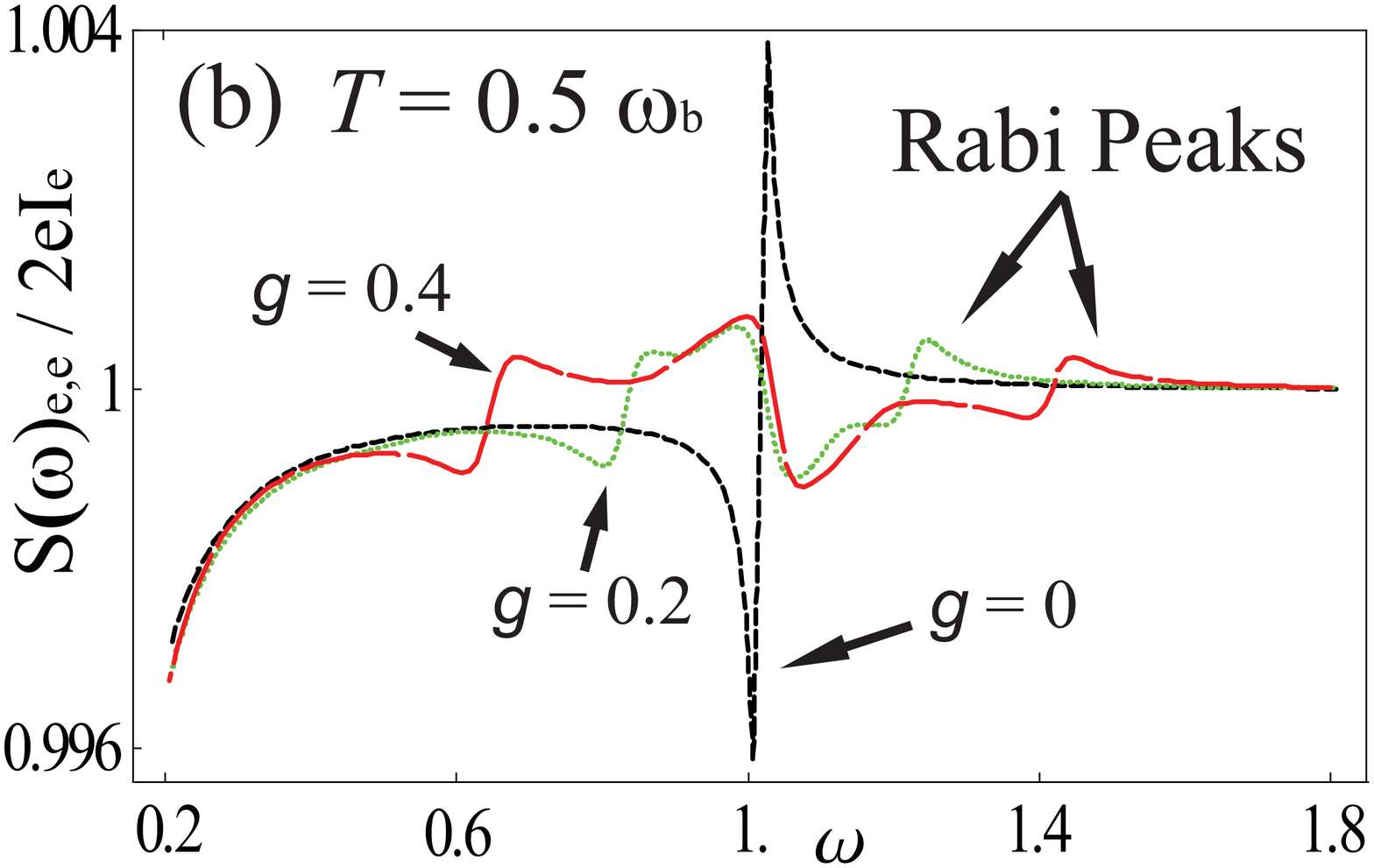}
\vspace{0.0cm}
\includegraphics[width=3.3in]{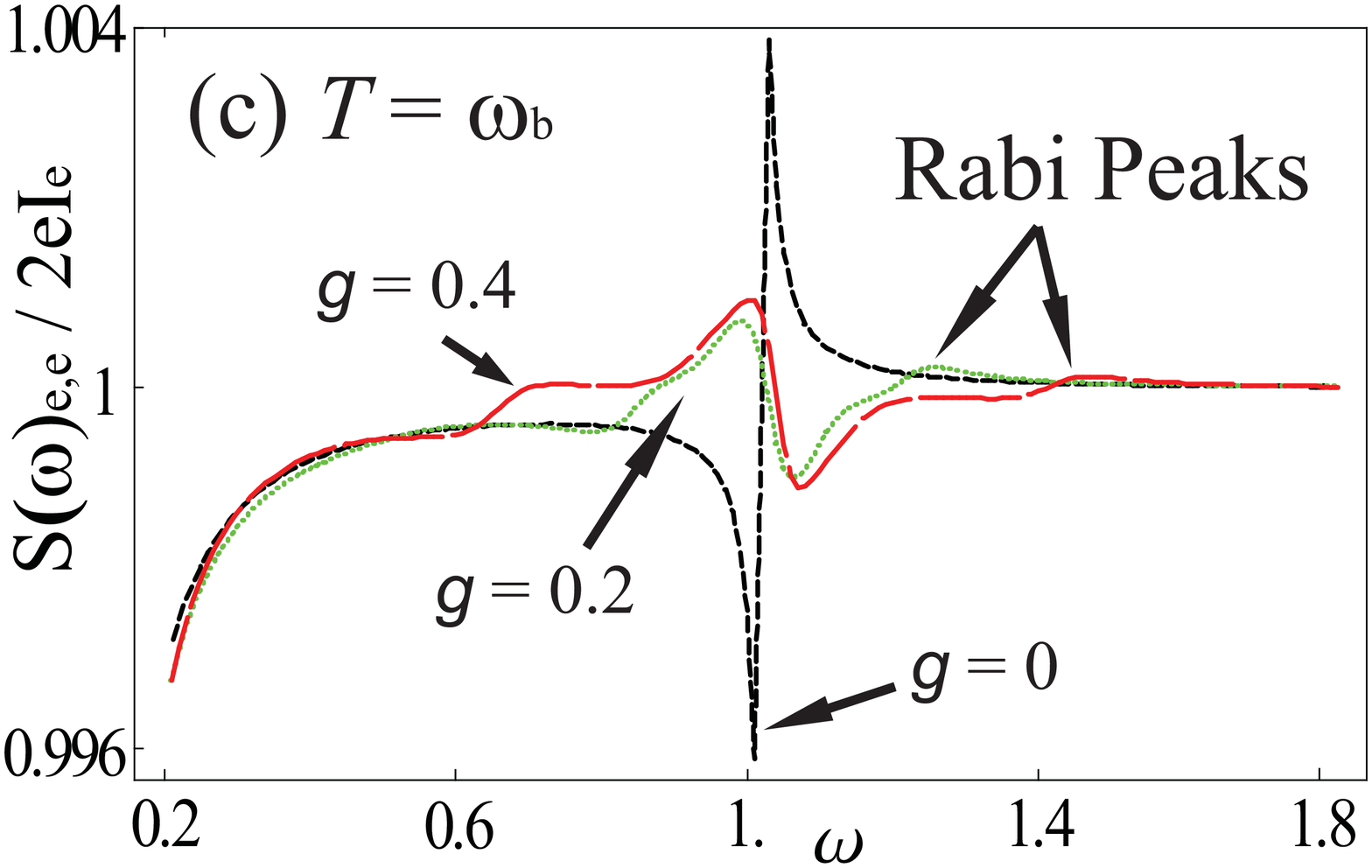}
\vspace{0.0cm} \caption{(color online) Normalized current-noise
frequency-spectrum, $S(\omega)_{e,e}/2eI_e$, versus $\omega$, for
$\omega_b=1$, $\Gamma_L=\Gamma_R=0.01$, $\Delta=0.5$,
$\gamma_b=0.05$, and $T=0$, $0.5\omega_b$, $\omega_b$ (figures (a),
(b), and (c), respectively) and a selection of electron-resonator
coupling strengths $g$; black, green, red are $g=0,0.2,0.4,$
respectively.
 The resonances are marked by arrows.
 As the temperature is increased, the system enters the classical regime and the two Rabi peaks decrease and become hard to distinguish. }\label{RWA2}
\end{figure}

\subsection{First regime:  effective Jaynes-Cummings Hamiltonian}

An effective Jaynes-Cummings Hamiltonian can be realized if we set
$\epsilon=0$ and $2\Delta=\omega_b$. Then, \beq H= \Delta\sigma_x+
g\sigma_z(a+a^{\dagger}) + \omega_b a^{\dagger}a . \label{2}\eeq
Large $\Delta$ implies~\cite{Brandes} that there is a strong overlap
between the particle wave functions in the two states, which may
introduce extra coupling terms with the resonator.  However, for
simplicity, we assume they are negligible.

First, we write the diagonal energy term for the qubit ($\sigma_z$)
in the off-diagonal basis ($\sigma_x$) by substituting raising and
lowering operators in that basis \beq \sigma_x^+ =
\frac{1}{2}(\sigma_z - i\sigma_y), \quad \sigma_x^{-} =
\frac{1}{2}(\sigma_z + i\sigma_y).\eeq  Then performing the
rotating-wave approximation in this basis, by dropping
counter-rotating terms, we obtain \beq \label{3} H_{JC}&\approx&
g(\sigma_x^+a+\sigma_x^-a^{\dagger}) + \omega_b
a^{\dagger}a + \Delta\sigma_x\\
&=&g\left[\frac{1}{2}(\sigma_z -
i\sigma_y)a+\frac{1}{2}(\sigma_z+i\sigma_y)a^{\dagger}\right] +
\omega_b a^{\dagger}a + \Delta\sigma_x. \nonumber\eeq  This has the
spectrum of an infinite number of non-interacting multiplets with
eigenstates, \beq \ket{\pm}_n=\frac{1}{\sqrt{2}}(\ket{n,1_x} \pm
\ket{n+1,0_x},\eeq where $\ket{n}$ is the number state of the
mechanical resonator, and $\ket{0}_x$ and $\ket{1}_x$ are the
eigenstates of $\sigma_x$ (i.e., the bonding and anti-bonding states
within the double quantum dot).

If we consider the zero-temperature limit and a strong damping of
the bath, then only the lowest number states of the mode $n=0,1$
strongly contribute to the transport processes (this case is well
into the quantum regime, and the ideal situation). This regime is
feasible if the effective temperature of the resonator is below
$\hbar \omega_b$. In this case, if $\Gamma_L \approx \gamma_b$ then
the initial state of each `round' of transport would be \beq
\ket{\psi(t=0)}=\ket{0,L}=\frac{1}{\sqrt{2}}(\ket{0,0_x} +
\ket{0,1_x}. \eeq 
The second component, $ \ket{0,1_x}$ couples to the $n=0$ and $n=1$
states of the mechanical resonator via the $\ket{\pm}_{n=0}$
eigenstates of $H_{JC}$.  The first component, $\ket{0,0_x}$, acts
as an `interaction free' transport route because it is the ground
state of $H_{JC}$. The component $\ket{0,0_x}$ has a unique `ground
state energy' $E_0=- \Delta$, while the two $\ket{\pm}_{n=0}$
eigenstates of $H_{JC}$
 have energies \beq E_{\pm}=\omega_b/2 \pm \sqrt{\Omega^2 + 4g^2}/2,\eeq where \beq \Omega=\omega_b -
 2\Delta.\eeq
Our numerical simulations in Fig.~\ref{RWA1} and Fig.~\ref{RWA2}
show clearly how the energy level {\em splittings} $\delta E_i$,
$(i=1,2,3)$ form {\em resonances} in the noise frequency spectrum.
In particular, because $\Omega=0$ here,  \beq \delta E_{1/2}=E_{\pm}
- E_0=2\Delta \pm g \eeq are the upper and lower resonance
`branches' in Figs. 2 and 3,  caused by the coherent coupling
between the double quantum dot and the mechanical resonator, and
$\delta E_3=2\Delta$ is the central resonance because of coherent
internal oscillations within the dot alone. This occurs because of
the $\ket{0,0_x}$ ground state of $H_{JC}$, described above, which
only evolves in time with a phase factor $E_0=-\Delta$.

As we increase the temperature of the mechanical resonator thermal
bath, the upper and lower resonance branches gradually disappear,
and the central resonance, determined by $\omega=2\Delta$,
dominates. Increasing the temperature of the `bath' means that the
mechanical resonator would be in a thermal mixture of number states;
thus for the electron, more transport channels become available.
This is more clearly apparent in the magnitude of the noise shown in
Fig.~\ref{RWA2}, illustrating that by {\em monitoring the peaks} in
the current-noise transport $S_{e,e}(\omega)$ one can, in principle,
{\em distinguish classical and quantum behavior.}  However, the
observation of near zero-temperature oscillations is not always
proof of quantum behaviour \cite{weig, blais,Schu08, Omelyanchouk,
Shevchenko,Franco1} as they can also be described by a classical
model of coupled linear oscillators. For example, in our
current-noise formulation ``false signatures'' from interactions of
the qubit with nearby classical oscillators may appear in the
spectrum and be mistaken for quantum Rabi behavior.  We discuss this
further in the next section.

One can understand the transition to the high-temperature case by
assuming the initial state to be \beq \ket{\psi(t=0)}=\left(\sum_n
C_n
\ket{n}\right)\otimes\frac{1}{\sqrt{2}}\left(\ket{0_x}+\ket{1_x}\right)\eeq
which connects each multiplet in the spectrum of the Jaynes-Cummings
Hamiltonian with its two nearest energy levels. The subspace of the
Hamiltonian connecting $\ket{n-1,1_x}$,$\ket{n,0_x}$ $\ket{n,1_x}$
and $\ket{n+1,0_x}$ is (where the basis here is for $\sigma_x$
diagonal),
\begin{widetext}\beq H_{n-1,n,n+1}=\left(
  \begin{array}{cccc}
        (n-1)\omega_b +\Delta & g\sqrt{n} & 0 & 0 \\
    g\sqrt{n} & n\omega_b -\Delta  & 0 & 0 \\
    0 & 0 & n\omega_b+\Delta & g\sqrt{n+1} \\
    0 & 0 & g\sqrt{n+1} & (n+1)\omega -\Delta \\
  \end{array}
\right) \eeq\end{widetext}
 Then, we easily see that the probability that the left dot is
 occupied (corresponding to the probability of the superposition of bonding and antibonding states
$\frac{1}{\sqrt{2}}(\ket{0_x}+\ket{1_x})$), is given by, \beq
P_L(t)=\sum_{n=0}^{\infty} \left\{C_n \cos
\left[\frac{-g(\sqrt{n+1}-\sqrt{n})t - 2t\Delta
}{2}\right]\right\}^2 \label{PL}\eeq Considering both an equal
superposition, $C_n=1/\sqrt{N}$ (but with cut-off of the sum in
$P_L(t)$ at a given $N$), and a coherent state distribution,
$C_n(z)=z^ne^{-z}/n!$, we observe that the oscillations in the the
probability $P_L(t)$ collapse over time, until only small
oscillations with period $\Delta$ around $P_L=0.5$ remain. This is
because the non-commensurate Rabi frequencies in Eq.~[\ref{PL}]
interfere {\em destructively}. For a small number of number states
$N$, or a small coherent state distribution $z$, there is some
revival in $P_L(t)$, but as $N$ increases the number of revivals
fall. This is also true if the initial state is a separable density
matrix with the resonator state in a thermal Boltzman distribution,
as is the case for high-temperatures.

\subsection{Second Regime:  Off-resonant interaction}
\begin{figure}[htp]
\centering \vspace{0.0cm}
\includegraphics[width=3.3in]{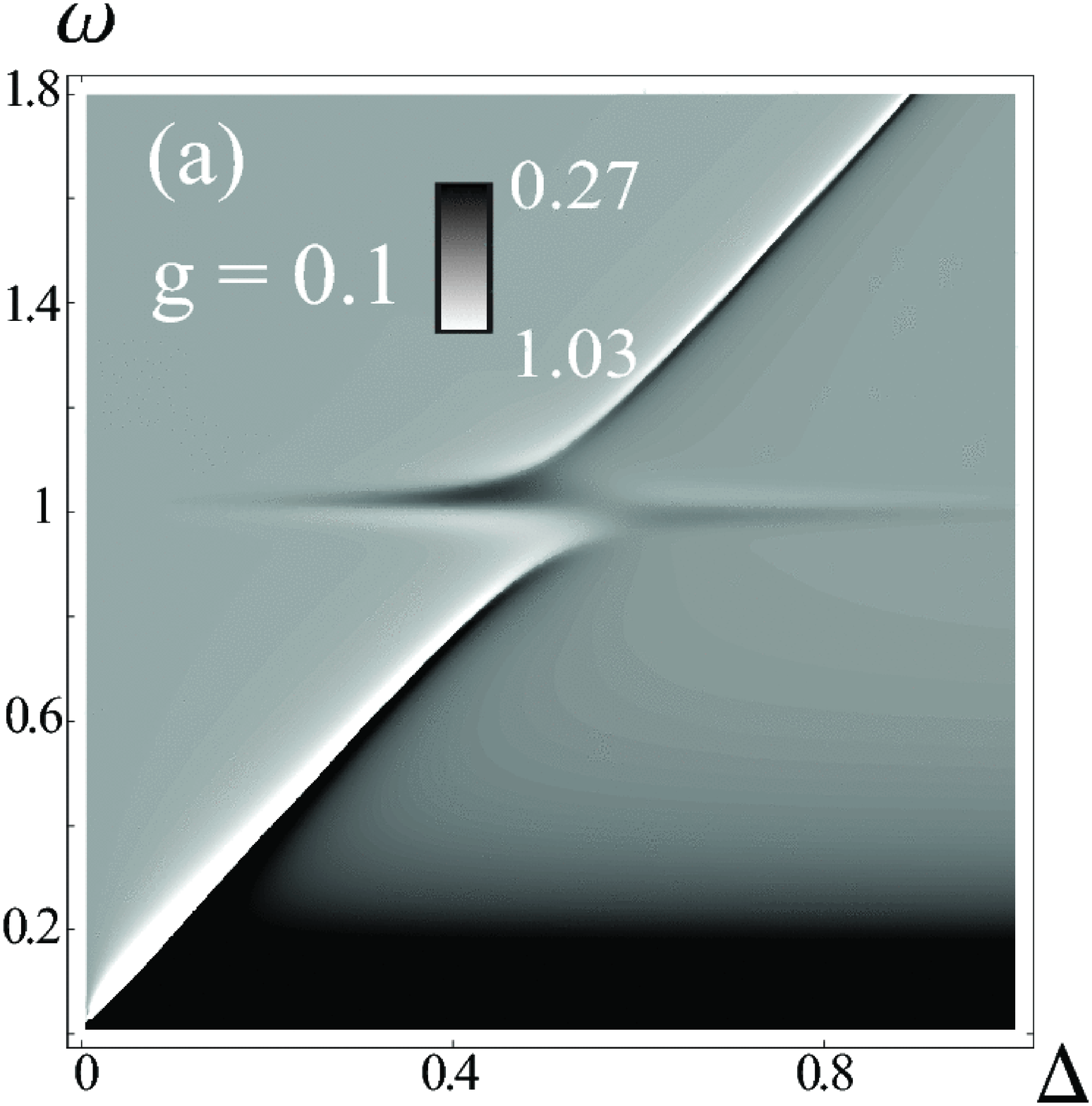}
\vspace{0.0cm}
\includegraphics[width=3.3in]{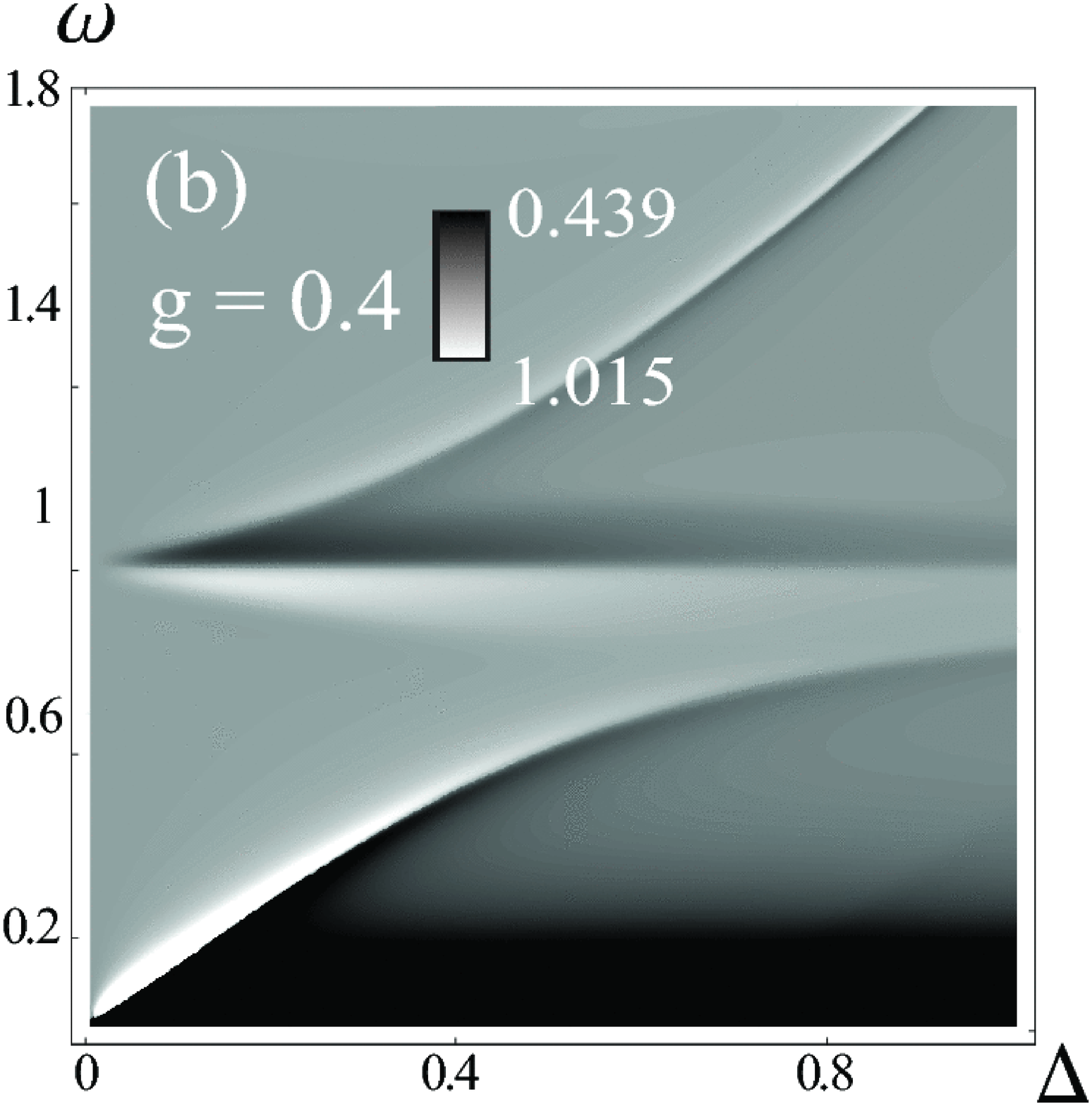}
\vspace{0.0cm}
\caption{  Current-noise frequency-spectrum,
$S(\omega)_{e,e}/2eI_e$, versus both the tunneling rate $\Delta$ and
frequency $\omega$, for (a) $g=0.1$ and (b) $g=0.4$. In both cases
$\omega_b=1$, $\Gamma_L=\Gamma_R=0.01$, $\gamma_b=0.05$,
$\epsilon=0$, and $T=0$. This was obtained numerically by solving
the master equation in Eq. (3). As $\Delta$ approaches
$\omega_b/2=0.5$ we see the three resonance points previously shown
in Fig. 3. Increasing $g$ increases the gap between the resonant
peaks. The hyperbolic behavior
 comes from the well know Jaynes-Cummings eigenvalue spectrum, as recently observed experimentally in a similar system (e.g., Refs~[\onlinecite{Deppe,weig,blais}]).} 
\label{nonrwa}
\end{figure}

In the previous section we showed that on-resonance,
$2\Delta=\omega_b$, the lowest part of the energy spectrum of the
coupled system was visible in the current-noise.   We can now verify
that these resonances really stem from the low-energy spectrum of
the Jaynes-Cummings Hamiltonian, and indicate coherent {\em quantum}
dynamics, by inspecting the {\em off-resonant} regime, $2\Delta \neq
\omega_b$, where the energy levels have a hyperbolic behavior.
In terms of the double-dot realization, we point out that assuming a
small $\Delta$ implies a tight confinement of the electron within
each dot.

Observing Fig.~4(a), we can see upper and lower resonance branches,
but in this case ($2\Delta \neq \omega_b$) they have the typical
hyperbolic tails of an avoided level crossing. In addition, Fig.
\ref{nonrwa}(b) shows that, as the coupling to the mechanical
resonator $g$ is increased, the gap in the level crossing increases.
Once more we are successfully observing the low-energy spectrum of
the Hamiltonian in the power spectrum of the current-noise. For
example, the upper and lower branches are simply given by the lowest
eigenvalues of the Jaynes-Cummings Hamiltonian, \beq \delta E_{1,2}
&=& |\omega_b/2 \pm \sqrt{\Omega^2 + 4g^2}/2+\Delta|.\nonumber \eeq

Furthermore, we note that there is an energy gap which halts the
electron current in the limit when $\epsilon=0$ and when the
coherent tunneling within the dots is small relative to the coupling
to the mode, $\Delta\ll g$. This occurs because the tunneling of an
electron requires an energy loss proportional to the displacement of
the mode, and because the rotating wave approximation is no longer
valid. For transport to occur, the electron must tunnel from the
left to the right state, which is now shifted in (relative) energy
by $2g(a+a^{\dagger})$. This becomes more and more difficult as the
coupling $g$ is increased, resulting in a ``current blockade''
effect.

Finally, as discussed in the previous section,  we point out that
oscillations alone may not provide sufficient proof of quantum
behavior. Recent circuit-QED experiments\cite{blais,weig} have
focused on the idea of observing the square-root dependence of the
energy of the Jaynes-Cummings system on the photon occupation number
$n$, which is sufficiently distinct from the behavior seen in
classical models. However, the preparation of arbitrary Fock states
in a nano-mechanical resonator is not readily realizable at this
point in time.

\subsection{Zero frequency noise: comparing the single and
many-mode cases}

In the previous sections we showed how the low-energy levels of a
Jaynes-Cummings Hamiltonian can be seen in the {\em full}-frequency
current-noise spectrum.  However, most recent experiments have
focused on the {\em zero}-frequency noise.
  For example, Kie{\ss}lich et al \cite{kiesslich} showed, by
comparing experiment and theory, that coherent oscillations in a
double quantum dot produced {\em super}-Poissonian
$[S(0)_{e,e}/2eI_e
> 1]$ signatures in the zero-frequency noise, while incoherent
transitions (sequential tunneling induced by increasing the
temperature of the phonon bath) produce {\em sub}-Poissonian noise,
$[S(0)_{e,e}/2eI_e < 1]$.

\begin{figure}[htp]
\centering
\vspace{0.0cm}
\includegraphics[width=3.3in]{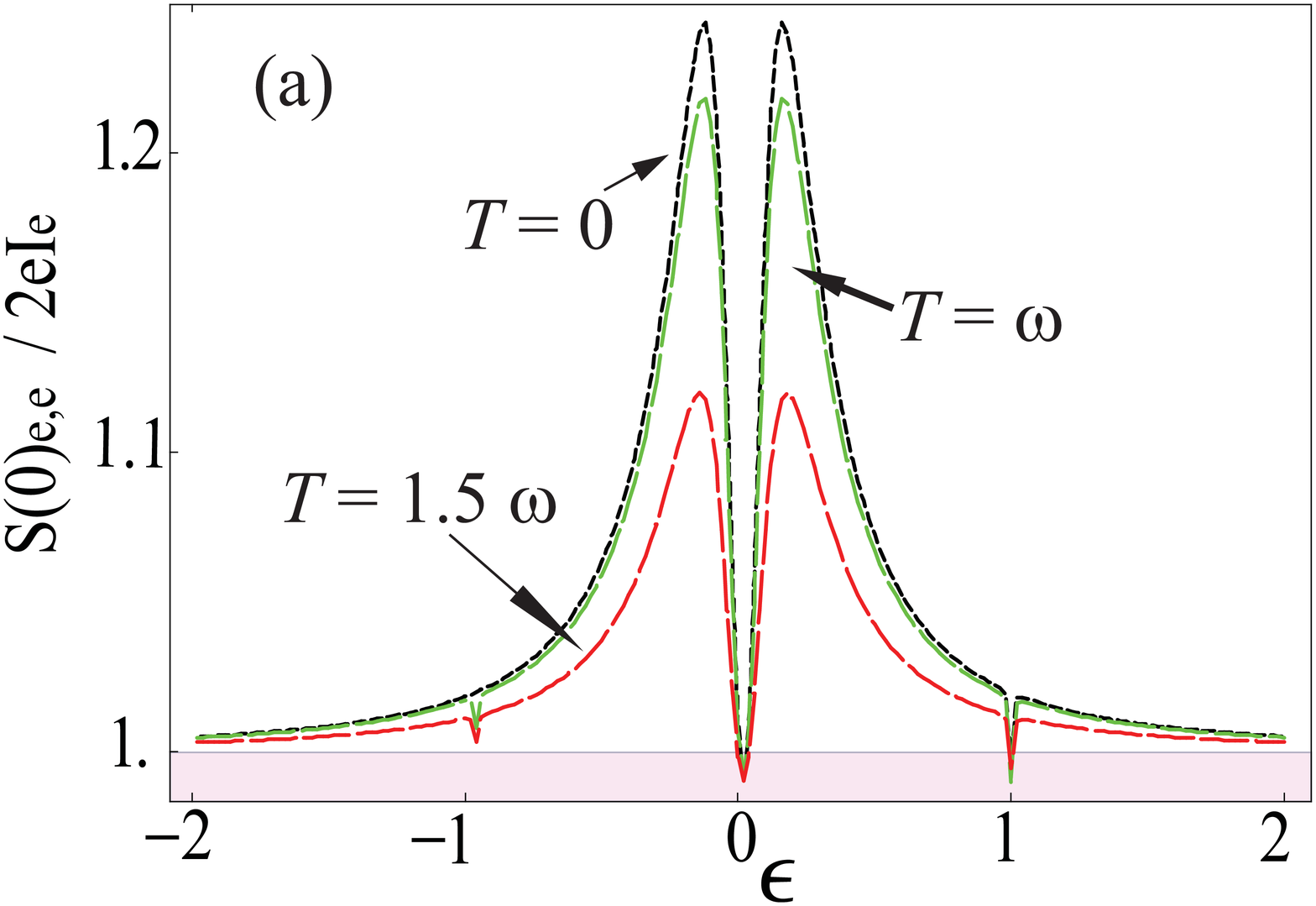}
\vspace{0.0cm}
\includegraphics[width=3.3in]{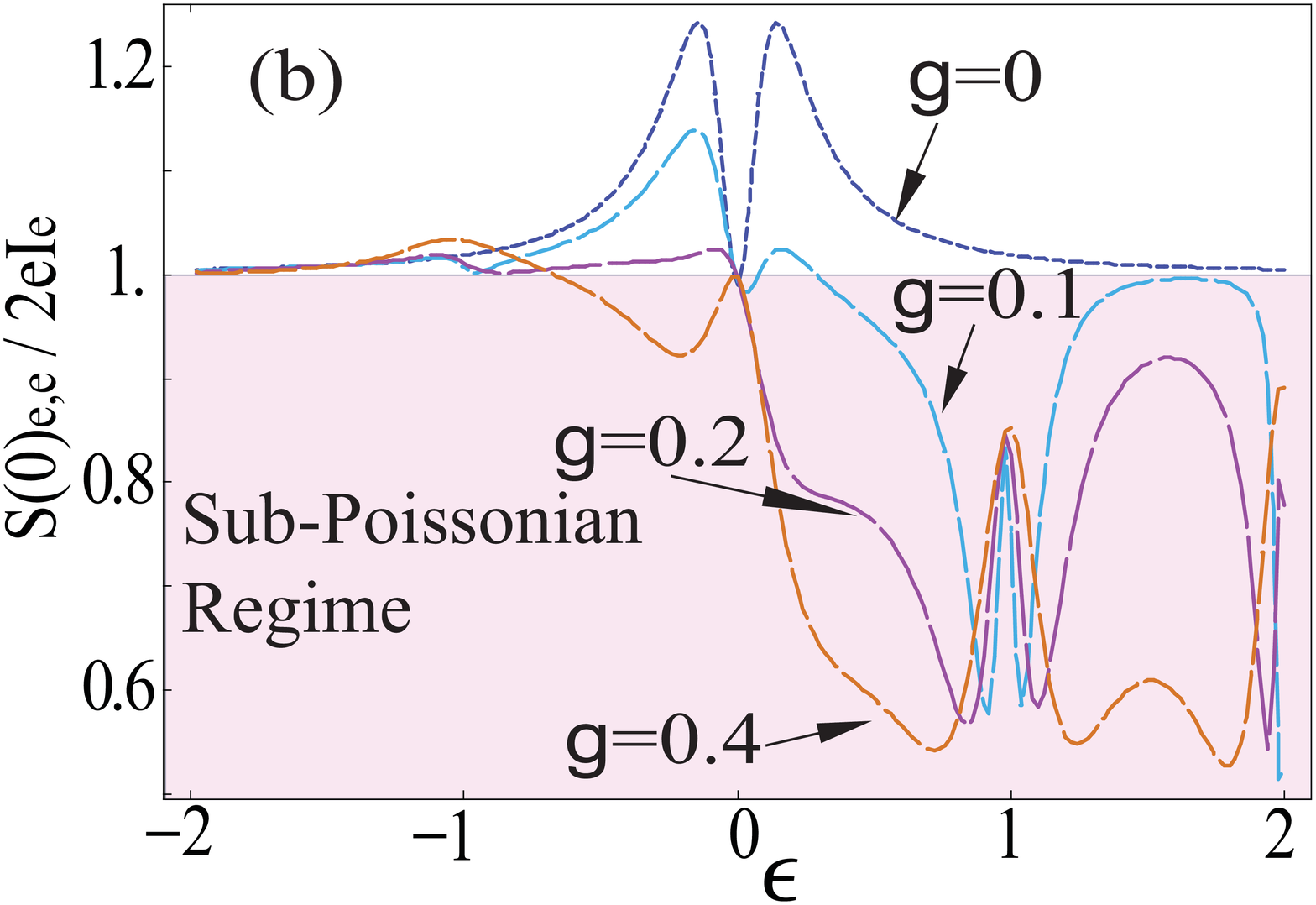}
\vspace{0.0cm}
\includegraphics[width=3.3in]{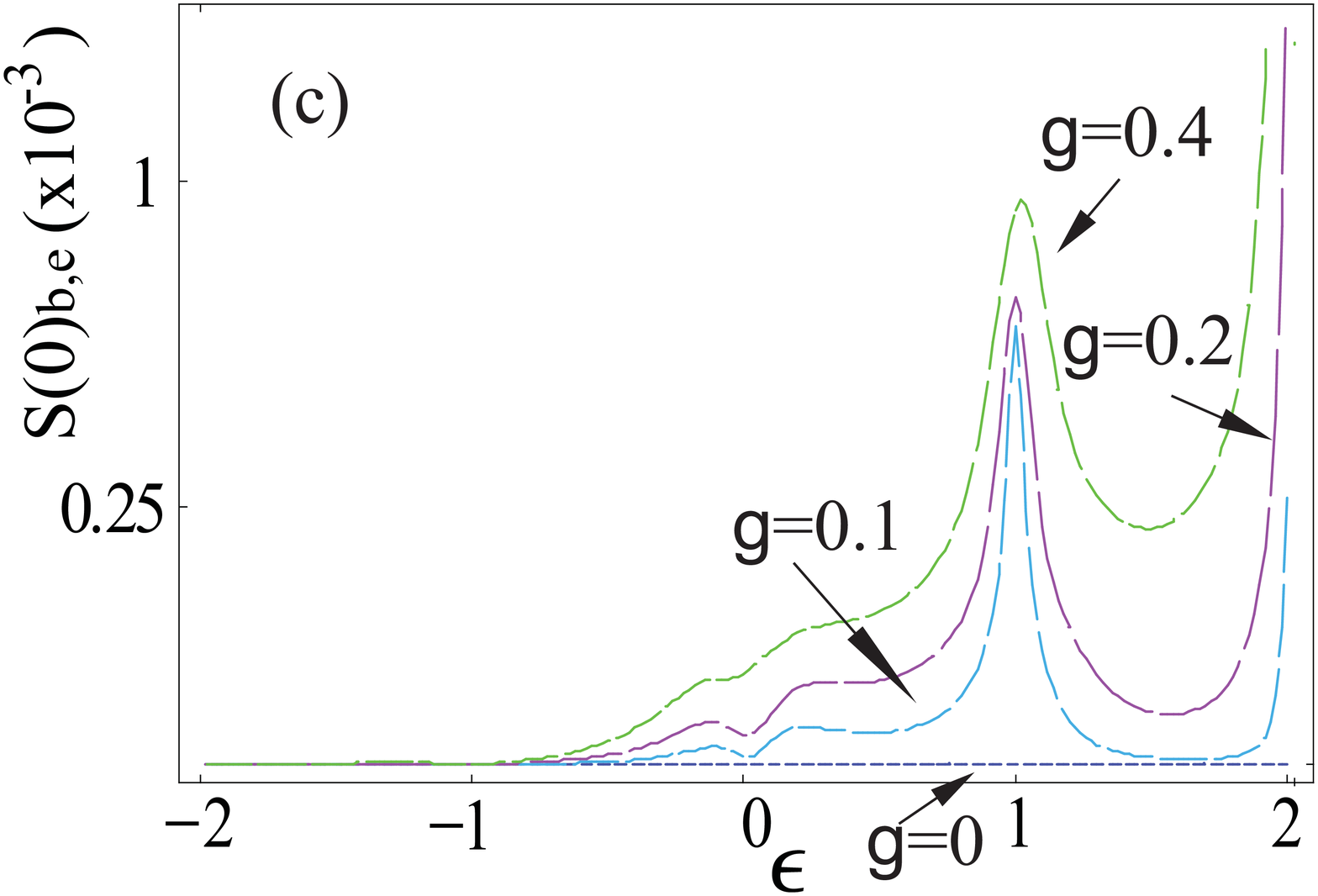}
\vspace{0.0cm} \caption{(Color Online) (a) Current-noise
frequency-spectrum $S(0)_{e,e}/2eI_e$ versus $\epsilon$, for
$\omega=1$, $\Gamma_L=0.1$, $\Gamma_R=0.001$, $\Delta=0.1$,
$\gamma_b=0.01$, $g=0.0008$, increasing $T$ from $0$ to $2$ in steps
of $0.5$. Increasing the temperature reduces the super-Poissonian
character of the current-noise.  (b) $S(0)_{e,e}/2eI_e$ for
$\omega=1$, $\Gamma_L=0.1$, $\Gamma_R=0.001$, $\Delta=0.1$,
$\gamma_b=0.01$, $T=0$, for $g$ from $0$ to $0.4$ in steps of $0.1$
(besides $g=0.3$, which has been omitted for figure clarity). For
strong coupling we see that the current-noise becomes almost
entirely sub-Poissonian (as indicated by the shaded pink region).
(c) The cross-correlation $S(0)_{e,b}$ against $\epsilon$ for the
same parameters as (b). Recall that $S(0)_{e,b}$ is defined as the
correlation between the electron and phonon currents. For $g=0$ and
for negative $\epsilon$ there is zero correlation between phonon and
electron tunneling, as expected.
 }\label{figs0}
\end{figure}

Mimicking their parameter regime, i.e. considering their device as
coupled to a resonator (or phonon cavity), now we also look at the
zero-frequency noise (as a function of double quantum dot level
detuning $\epsilon$). We observe similar signatures to theirs in the
noise spectrum, but with a more complicated structure.  
  We also observe, in Fig.\ref{figs0}(a), that increasing the temperature of the {\em single
mode} resonator decreases the zero frequency current-noise,
eventually resulting in sub-Poissonian behavior.

Similarly, increasing the bare coupling strength $g$ to the single
mode resonator has a drastic effect.   As Fig. \ref{figs0}(b) shows,
the noise profile quickly becomes sub-Poissonian, developing a new
peak structure around $\epsilon=1$. Interestingly, the $\epsilon=0$
point, where we earlier probed for coherent signatures, remains
around $S(0)_{e,e}/2eI_e=1$, indicating that coherent transport is
still occurring.

Figure \ref{figs0}(c) illustrates the (non-normalized)
cross-correlated noise, i.e. the correlation between electron
tunneling events and phonons leaving the mechanical resonator into a
heat bath (with rate $\gamma$). As expected, there is no correlation
between tunneling events when the systems are uncoupled.
Furthermore, the correlated noise is large when $\epsilon=
k\omega_b$, where $k$ is an integer. While the correlated noise
grows for larger `$k$', the current itself becomes
smaller~\cite{Lambert03}. This is simply because as $\epsilon$
increases, the current can only flow through phonon assisted
tunneling, which happens at integer numbers of the phonon frequency.

\section{Squeezing the quantum state of the resonator}

We have shown that the electron current-noise, $S_{e,e}(\omega)$,
serves as a detector of coherent interactions between the double
quantum dot and the single mode of the mechanical resonator. Already
this is a significant step, as $S_{e,e}(\omega)$, serves as a tool
for experimental observation. However, we can proceed a step
further, and briefly consider the statistics of the phonons in the
mechanical resonator. In such a mechanical system, these quantities
are difficult, if not impossible, to access. However, it is
informative to understand how the phonon statistics of the resonator
change as we increase the temperature, and leave the quantum regime.
\subsection{Squeezing signatures}
In the proposal by Rodrigues et al \cite{Rodrigues} they show that
the resonator can exhibit properties akin to a micromaser, due to
the nonlinear coupling to an SSET.  In their case, the qubit is
represented by  a superposition of island charge states
$\sigma_z=\ket{2}\bra{2}-\ket{0}\bra{0}$. However, they focused on
the regime where $\omega_b/\Gamma=1$, observing that this is where
the interaction between the resonator and SSET is maximized. In the
results we have shown in the previous sections, we assume that the
quantum dots and leads are weakly coupled, $\omega_b/\Gamma \gg 1$.
Furthermore, we assume that the resonator is strongly damped (e.g.,
via cooling by another SSET, or by the double quantum dot itself
\cite{Franco3, Ouyang}), so that only the few lowest bosonic levels
are excited.

However, even for our `slow' regime, we see sub-Poissonian
signatures in the {\em boson emission} noise spectrum emitted into
its nearby heat bath $S_{b,b}(0)/2I_b$, as well as in the Fano
factor $F_Q$ of the number state occupation $n$ of the
resonator\cite{Hu96,Hu962,Hu97,Hu99} \beq
F_Q=\frac{\ex{n^2}-\ex{n}^2}{\ex{n}},\eeq as shown in Fig. 6(a) and
(b). Both `measures' identify similar regions of squeezing, though
there is a conceptual difference between the squeezing of the
phonons emitted (dynamically) into the heat bath, and a direct
measurement of the static steady-state phonon occupation number.
Furthermore, we see that as the temperature is increased, both
quantities increase non-linearly in magnitude.

In addition, we consider the correlated electron-phonon noise. We
naively expect that stronger correlations will occur in the quantum
regime.   Figure 6(c) verifies this, and shows that a maximum in the
correlated noise occurs around $g=0.1$, and an increase in
temperature reduces the overall magnitude. This is an indication,
continuing from previous suggestive results \cite{Lambert07}, that
the quantum noise correlation between two open systems could serve
as a measure of entanglement, though a direct correspondence has yet
to be identified.

\subsection{Quadrature versus number-state squeezing}

The squeezing in Fig. 6 is {\em number state squeezing}, and a
sub-Poissonian variance in $n$ ($F_Q$) implies anti-bunching of the
phonon statistics\cite{teich}.   This is only one of several types
of squeezing. For example, in quantum optics, generalized quadrature
squeezing is often investigated.  Typically the axis of squeezing
might not been known, so a homodyne measurement of the occupation
statistics must be performed.  A homodyne measurement \cite{Mandel},
using a local oscillator to introduce a relative phase, reveals the
variance of any desired quadrature.    Thus, in principle, it is
possible to measure the normal ordered squeezing via \beq
\ex{:(\Delta Q)^2:}=\ex{:Q^2:}-\ex{Q}^2, \eeq where $Q$ is the
quadrature defined by a desired angle $\phi$, so that \beq Q=a e^{-i
\phi} + a^{\dagger}e^{i\phi}. \eeq Squeezing of the quadrature is
implied when \beq \ex{:(\Delta Q)^2:} < 0 \eeq for some given
$\phi$, because of the normal ordering. Again, in a nanomechanical
system such a measurement is not feasible, but has been proposed in
transmission line resonators\cite{Lehnert}. Is is trivial to see
\beq \ex{:(\Delta Q)^2:}&=&
\ex{a^{\dagger^2}}e^{2i\phi}+\ex{a^2}e^{-2i\phi}
+2\ex{a^{\dagger}a}\nonumber \\ &-&
\ex{a^{\dagger}}^2e^{2i\phi}+\ex{a}^2e^{-2i\phi} +2\ex{a}\ex{
a^{\dagger}}. \eeq  However, for our model and parameter space,  we
were not able to observe any instance of quadrature squeezing.  In
the previous sections we discussed how strong contributions to the
steady-state solution of the master equation arise from the
low-level Jaynes-Cummings eigenstates.   Our results illustrate
that, in our system, these states only produce number state
squeezing in the resonator mode, but not quadrature
squeezing\cite{teich}.

\begin{figure}[htp]
\vspace{0.0cm} \centering
\includegraphics[width=3.3in]{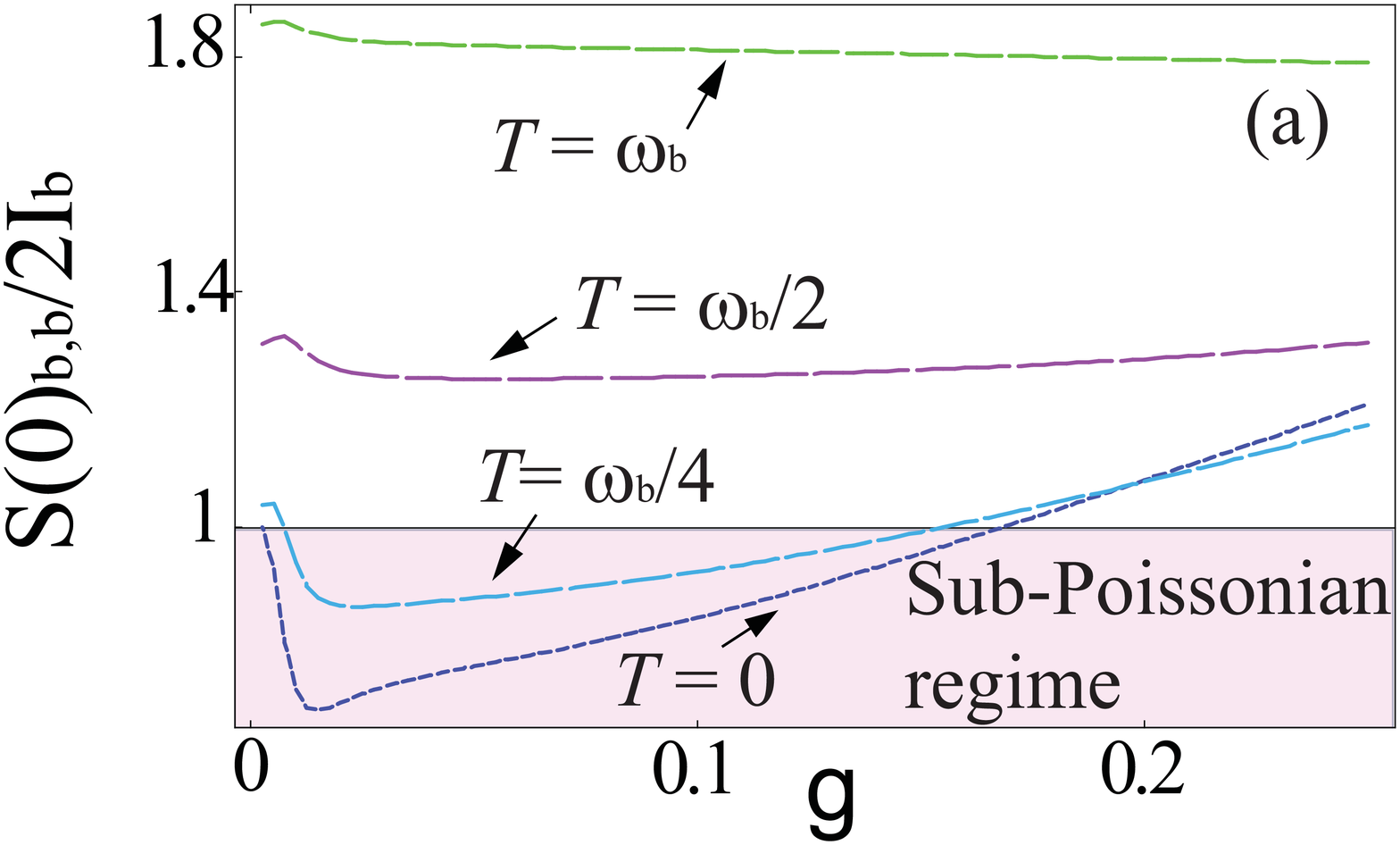}
\vspace{0.0cm}
\includegraphics[width=3.3in]{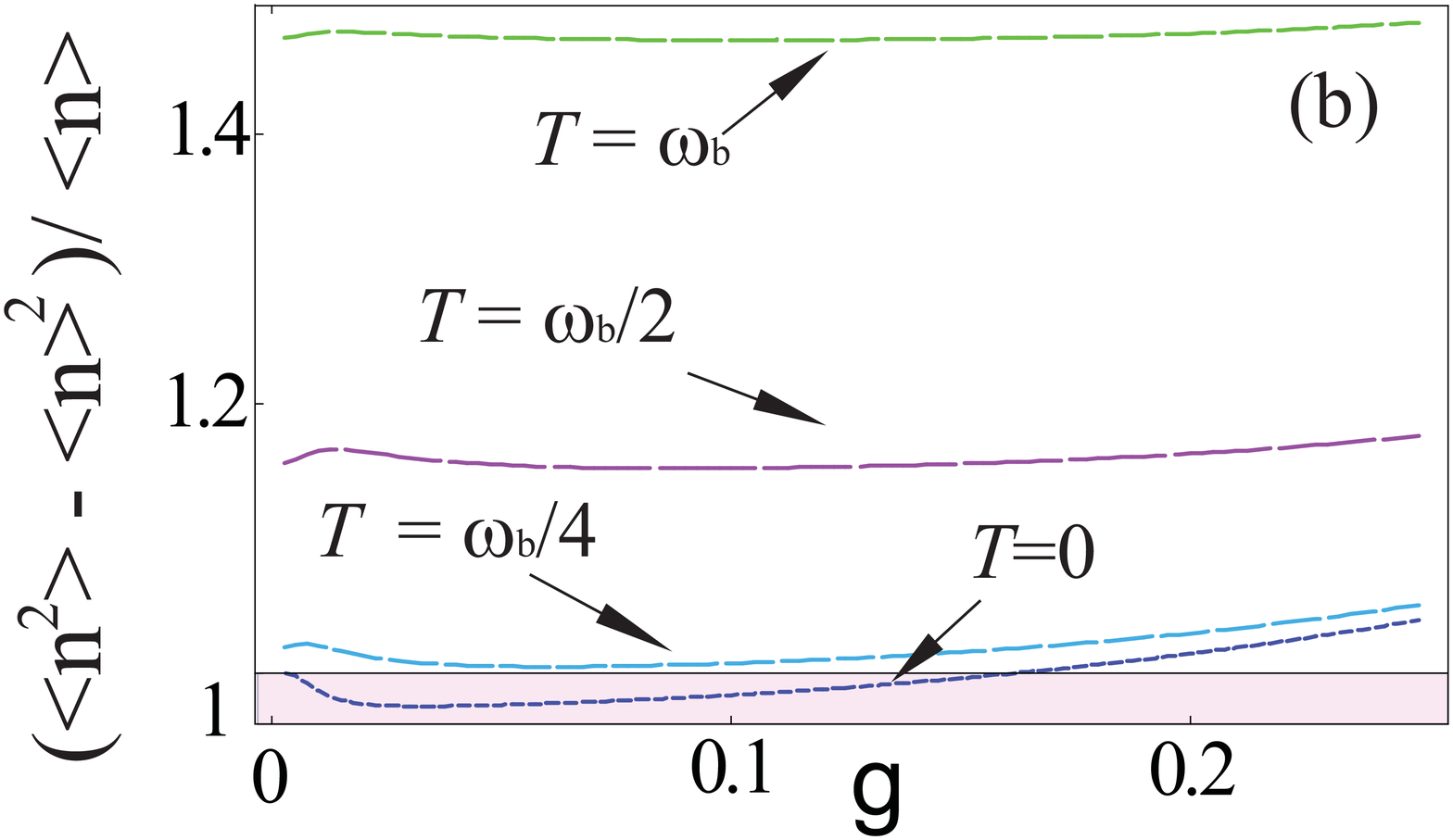}
\vspace{0.0cm}
\includegraphics[width=3.3in]{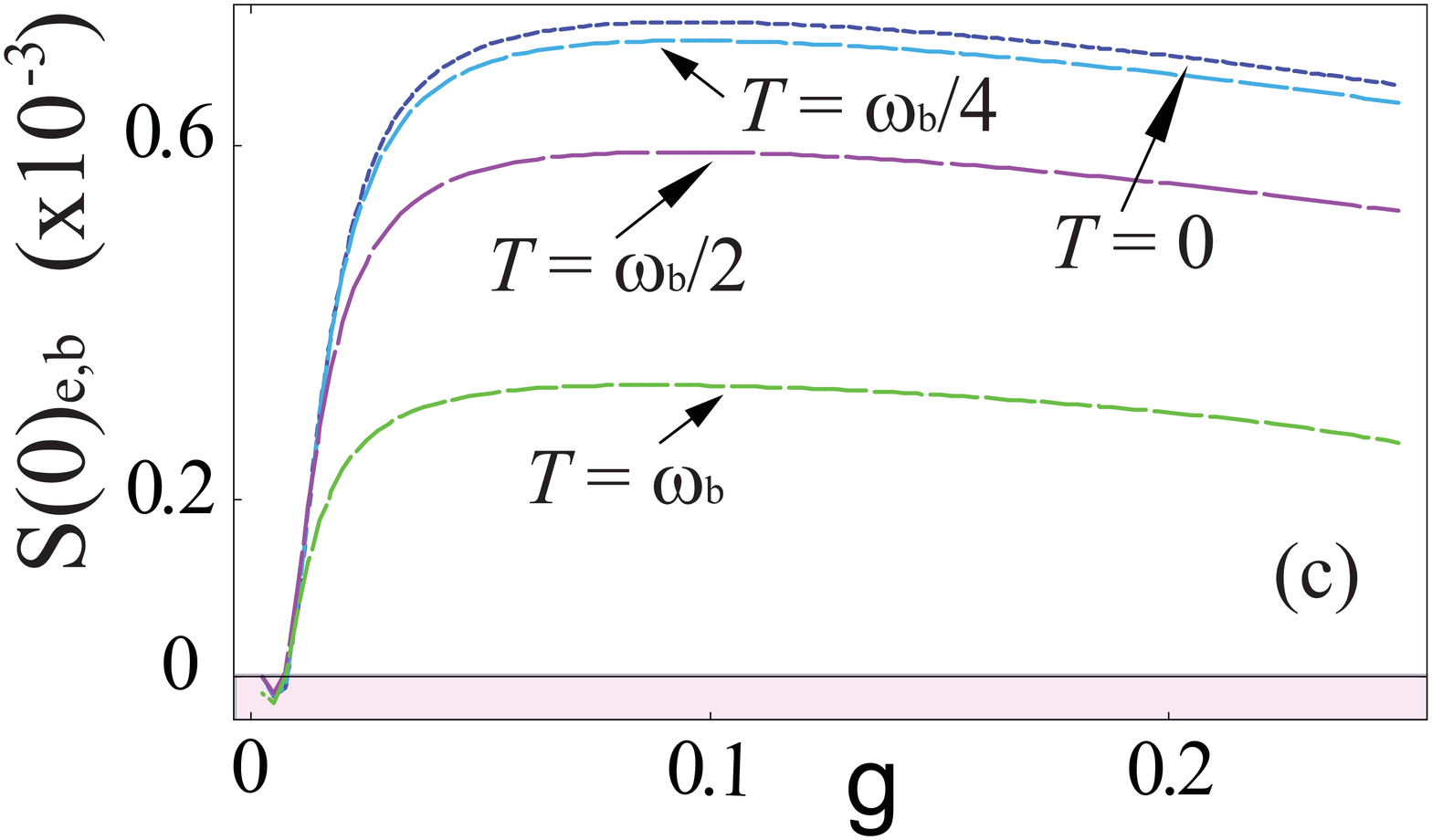}
\vspace{0.0cm}\caption{(Color online) (a) Bosonic current-noise
frequency-spectrum, $S_{b,b}(\omega)/2I_b$, versus $g$, for
$\omega_b=1$, $\Gamma_L=\Gamma_R=0.01$, $\Delta=0.5$,
$\gamma_b=0.05$, $\epsilon=0.0$, and a range of temperatures. The
pink regions [also denoted by "Sub-Poissonian regime" in (a)]
indicate the regimes where quantum state squeezing occurs, for low
temperature and intermediate couplings. (b) shows the number-state
Fano factor $(\ex{n^2}-\ex{n}^2)/\ex{n}$ of the bosonic system. The
zero-temperature case closely corresponds to the `phonon
current-noise' in (a).  (c) shows the electron-phonon correlated
noise $S_{e,b}(0)$ versus $g$. Interestingly, {\em increasing} the
temperature {\em decreases} the {\em correlated} noise.  Moreover,
and as expected, the zero coupling point ($g=0$) remains around
$S(0)_{e,b}=0$ for all values of temperature.   Also, in a very
small regime of weak coupling $(g\rightarrow 0$), $S(0)_{e,b}$ can
be negative. }
\end{figure}
\section{Realizations}

As mentioned before, our model can correspond to charge states in a
double quantum dot in a  capacitively-coupled or suspended geometry.
For the suspended geometry\cite{Lambert03, Blich}, it has been shown
that there is a direct coupling between the electron wave function
and a single phonon mode because of van-Hove singularities in the
density of states. However such experiments have not yet been
performed in the energy regime of the fundamental {\em vibrational}
mode of a mechanical resonator. Also, our model is related to that
of a superconducting single electron transistor (SSET) capacitively
coupled to the resonator \cite{Schwab2, Rodrigues,tobias04}.
Typically there are some differences in the transport properties as
an SSET is a three-terminal device, and
 the SSET drives the resonator into complex types of limit-cycle
behavior \cite{Rodrigues,Armor08}.

\subsection{Energy scales}
To check the feasibility of our results we need to verify the
appropriate energy scales in real systems. We assume that our
state-of-the-art resonator has a fundamental frequency of
$\omega_b=1$ GHz. The corresponding `resonant' bias, $\epsilon=\hbar
\omega$, is approximately $4$ $\mu$eV. We assume we are near the
quantum limit, i.e., $kT\approx \hbar \omega_b$, $T\approx 50$ mK.
Normal capacitive coupling strengths for an SSET are $100$ MHz,
corresponding to $g=\omega/10$. In Figure 2 we saw signatures of
quantum coherent oscillations for this range of coupling strengths.
The same range ($g\sim\omega/10$) is feasible for the coupling
between a double quantum dot and the resonator (with capacitive
coupling~\cite{Ouyang}). The achievable coupling strengths for
suspended geometries are not precisely known now, but because of
van-Hove singularities in the density of states one can expect large
effective coupling strengths \cite{Lambert03, Blich}. Finally, the
inter-state tunneling, denoted by $\Delta$ in our discussion, is
typically tunable for double quantum dots. Thus a range of
$\Delta\sim (1$--$10)$ $\mu$eV is feasible.

\subsection{Magnetized resonator interacting with electron spins}
A recent proposal~\cite{Lambert08} focused on a magnetized resonator
which interacts with one of two electron spins in a spin-blockaded
double quantum dot system.  In this case, the current is used to
measure the spin state because, if the two spins are parallel,
current cannot flow.  An oscillating magnetic field, from the
magnetized resonator, couples to one of the spin states, and thus
this spin plays the role of a `transport qubit' in our earlier
language.  The question of cooling such a magnetized resonator and
then coupling it to a nearby electron spin via its quantized motion,
and henceforth the quantized magnetic field motion, has not been
addressed.  In that case, the Hamiltonian of the spin and the
resonator is, \beq \label{4} H_Q = -\,\frac{\Sigma\sigma_z}{2} +
\hbar \omega_b\, a^{\dagger}a +
C\sqrt{\frac{\hbar}{2m_{\mathrm{eff}}\omega_b}}(a+a^{\dagger})\sigma_x
\eeq where $C=0.16$ mT/nm.  This (Eq. \ref{4}) differs from the
Hamiltonians in Eqs. \ref{2} and \ref{3} in that Eq. \ref{4} is
diagonal in the qubit energy basis. The ground state motion of a $1$
GHz resonator is $2\times 10^{-14}$ m, which, using the parameters
from Ref. \onlinecite{Lambert08}, would generate a field of just
$3.2\times 10^{-6}$ mT, a Rabi frequency of about $100$ Hz, which is
negligible in comparison to nuclear hyperfine and spin-orbit
effects. Optimizing device design can increase this Rabi frequency
considerably. For example, a larger magnetization could be achieved
by using a Dysprosium (Dy) micromagnet instead of Cobalt (Co)
(giving a factor of about two). Similarly, a larger micromagnet
thickness could also contribute a factor of about two to the field
felt by the electron spin. Decreasing the distance between the dot
and resonator could contribute up to a factor of ten, and using a
slower frequency resonator, for a larger ground state displacement,
 could add a factor of about five.  Taking these factors into consideration
gives a Rabi frequency in the range $10$--$100$ kHz.  This Rabi
frequency is still, in comparison to the charge-based quantum dot
and SSET systems, a weak coupling, and is vulnerable to dephasing
from nuclear hyperfine fields. However, the future evolution of this
technology may make such an approach feasible and desirable,
especially considering the possible benefits of combining
spintronics and nanomechanics.

\section{Conclusions}

We have illustrated how quantum coherent behavior and the energy
spectrum of a nanomechanical resonator can be identified using
full-frequency current-noise measurements through a nearby transport
qubit.  In the zero-frequency limit, we showed that a single-mode
`environment', as represented by a nanomechanical resonator,
produces unique signatures that differ from those observed in
multi-mode environments.   Furthermore, we identified regimes where
phonon squeezing and cross-correlated noise, indications of complex
quantum phenomena, could occur.  All of these features could be
realized with a
 double quantum dot or superconducting single-electron transistor operating as the transport
 qubit.  In a broader context, we expect that noise measurements could also be
 useful in two-resonator circuit QED systems\cite{Mar05, sun, regal},
 which may offer an interesting area for future investigation.

\acknowledgements

We  thank Sahel Ashab, Christoph Bruder, Tobias Brandes, and
Yueh-nan Chen for helpful discussions.  FN acknowledges partial
support from the National Security Agency (NSA), Laboratory for
Physical Sciences (LPS), Army Research Office (ARO) and National
Science Foundation (NSF) grant No. EIA-0130383.

\appendix

\section{Noise Formalism}

To calculate the quantum noise~\cite{Blanter00} of a system with
Hamiltonian $H$, and corresponding transport environment described
by a Liouvillian $L$,
we employ a generating function approach.  
The Master equation for the matrix elements of the generating
function ${\bf g}$ is
\begin{eqnarray}
  \frac{\partial }{\partial t} {\bf g}(s_1,...,s_m,t) = M(s_1,...,s_m)\,{\bf g}(s_1,...,s_m,t),
\end{eqnarray}
which can be formally solved by diagonalizing \beq M(s_1,...,s_m)&
=& \\V(s_1,...,s_m) D(s_1,...,s_m) V^{-1}(s_1,...,s_m).
\nonumber\eeq Here $M$ is the Liouvillian $L$ recast as a function
of the counting variables $(s_1,...,s_m)$.  Each $s_i$ is a
continous variable which tracks the passage of the current through
system $i$. This gives a general formalism for calculating the
generating function of $m$ coherent and interacting transport
systems, each with a single `one-way' current flow.

 The next step is to use the MacDonald
formula \cite{Macdonald62} for the {\em symmetrized noise power
correlator} between systems $i$ and $j$ \beq S(\omega)_{i,j}&\equiv&
\int_{-\infty}^{\infty}d\tau e^{i\omega \tau}
\left[\ex{\delta{I_i(t+\tau)},\delta{I_j(t)}}\right]_{t\rightarrow\infty}\\
\frac{S(\omega)_{i,j}}{2e^2\omega }&=& \int_{0}^{\infty} \sin
(\omega \tau)\partial_{\tau} \left(\ex{n_i(\tau)n_j(\tau)} -
\frac{\tau^2 \ex{I_i}\ex{I_j}}{e^2}\right) ,\nonumber\eeq which can
be written as ($s=\{s_1,s_2,...,s_m\}$)
\begin{eqnarray}\label{Mc_Donald}
\frac{S(\omega)_{i,j}}{2e^2\omega }
&=&\left(\partial_{s_i,s_j}+\delta_{i,j}\partial_{s_i}\right)
\int_0^{\infty}d\tau
\sin(\omega \tau) \\
&\times&\frac{\partial}{\partial \tau}{\rm
Tr}\hat{G}(s\tau)|_{s=1},\nonumber
\end{eqnarray}
where an omitted term $2\tau \langle I \rangle^2$ in the integral
does not contribute in the final result obtained upon performing the
Laplace transformation.  Noting that $\hat{G}(s,\tau=0)=\rho(0)$,
where the initial condition $\rho(0)$ is the steady state density
matrix and using \beq \frac{\partial}{\partial \tau}\hat{G}(s,\tau)=
M(s) \hat{G}(s,\tau)=  M(s)e^{\tau
M(s)}\hat{G}(s,\tau=0)\nonumber,\eeq and the spectral decomposition
of $M(s)$, one obtains
\begin{eqnarray}\label{central_result}
  S(\omega)_{i,j} &=& 2e^2 \left(\partial_{s_i,s_j}+\delta_{i,j}\partial_{s_i}\right)\nonumber\\
&\times&\left\|  V(s) \frac{\omega^2 D(s)  }{\omega^2 +D(s)^2}
V^{-1}(s) {\bf g}(s,0) \right\|_{s=1},
\end{eqnarray}
where the notation $\left\|(x_{i_1,j_1},x_{i_2,j_2},...,)
\right\|\equiv \sum_{i=0}x_{ii}$ takes into account the trace in \
Eq.~(\ref{Mc_Donald}).  Note that the first derivative in the single
system correlator $\partial_s$ yields  $2e\langle I\rangle$, and
therefore  $\partial_s^2$ provides the deviation from the shot
noise.  Using the Ramo-Shockley theorem \cite{Blanter00}, the
displacement current contribution can either be omitted (by assuming
that the capacitances of the devices are extremely asymmetric, so
that $c_Lc_R\ll 1$), or calculated using a multi-variable approach,
because the total current fluctuations can be written as
\beq\label{equationI}
\delta I(t+\tau) \delta I(t)&=& \alpha^2 \delta I_L(t+\tau)\delta I_L(t+\tau) \nonumber\\&+& \beta^2 \delta I_R(t+\tau) \delta I_R(t)\nonumber\\
 &+& \alpha \beta (\delta I_L(t+\tau) \delta I_R(t)\nonumber\\ &+& \delta I_R(t+\tau)\delta
 I_L(t)).
\eeq The left and right correlations are trivially calculated using
separate counting variables for each lead.

Equation ~(\ref{central_result}){} allows one to calculate the noise
spectrum for transport through an arbitrarily complex quantum
system.  This can be evaluated either using finite difference
derivatives around $s=1$, or following the methods employed by
Flindt et al. \cite{Navotny04,Navotny042}.  In the latter case we
can use their approach to show that, in general, the
cross-correlator can be written as \beq
\partial_t  \ex{n_i(t)n_j(t)}  &=& \text{Tr}[ L_i \sum_{n_1,n_2,...} n_j  \, \rho^{(n_1),(n_2),...}]\nonumber\\
&+& \text{Tr}[L_j \sum_{n_1,n_2,...}n_i \, \rho^{(n_1),(n_2),...}].
\eeq Furthermore the terms  \beq \sum_{n_1,n_2,...} n_j
\rho^{(n_1),(n_2),...}=\partial_{s_j}\hat{G}(s,\tau)|_{s=1}\eeq can
be evaluated by Laplace transforming the equation of motion \beq
\partial_{\tau}\hat{G}(s,t)=(L_0 + \sum_i s_i L_i)\hat{G}(s,t)\eeq and
taking derivatives in the counting variables $s_i$, giving \beq
\partial_{s_i}\tilde{G}(s,-i\omega)|_{s=1}=F(-i\omega)L_iF(-i\omega)\rho(0)\eeq
where \beq F(-i\omega)=(-i\omega-L)^{-1}\eeq and $\rho(0)$ is the
steady-state initial condition.  As shown by Flindt et al
\cite{Navotny04} one can evaluate this inverse by writing \beq F(-i
\omega)&=&-P/i\omega - R(\omega),\\ R(\omega)&=&Q(i\omega
+L)^{-1}Q,\eeq where \beq P=\rho(0)\otimes 1,\quad Q=1-P.\eeq
Inserting all these expressions into the cross-correlator, and using
$P\rho(0)=\rho(0)$ and $Q\rho(0)=0$, gives the noise power as the
trace of an inverse, \beq
\frac{S(\omega)_{i,j}}{2e^2}&=&\text{Re}\left\{-\text{Tr}[L_iR(\omega)L_j\rho(0)]
-\text{Tr}[L_jR(\omega)L_i\rho(0)]\right\}\nonumber\\
&+&\delta_{i,j}\text{Tr}[L_i\rho(0)].
\eeq

All of the above allows us to calculate the full frequency spectrum
for an arbitrary number of coupled systems.  In addition, it allows
us to calculate {\em phonon} current and statistics.   We choose as
the phonon current operator the operator which absorbs a phonon
number state from the mode and puts it in the background bath.

\bibliography{bibliography}

\end{document}